\documentclass[12pt]{article}

\usepackage{amsmath,amssymb,amsfonts}
\usepackage{graphicx}

\makeatletter

\renewcommand\section{{\setcounter{equation}{0}}
			       \@startsection{section}{1}{\z@}
                                   {-3.5ex \@plus -1ex \@minus -.2ex}
                                   {2.3ex \@plus .2ex}
                                   {\normalfont\large\bfseries}}
\renewcommand\subsection{\@startsection{subsection}{2}{\z@}
                                   {-3.25ex\@plus -1ex \@minus -.2ex}
                                   {1.5ex \@plus .2ex}
                                   {\normalfont\normalsize\bfseries}}
\renewcommand\subsubsection{\@startsection{subsubsection}{3}{\z@}
                                   {-3.25ex\@plus -1ex \@minus -.2ex}
                                   {1.5ex \@plus .2ex}
                                   {\normalfont\normalsize\bfseries}}
\renewcommand\paragraph{\@startsection{paragraph}{4}{\z@}
                                   {3.25ex \@plus1ex \@minus.2ex}
                                   {-1em}
                                   {\normalfont\normalsize\bfseries}}

\makeatother

\newcommand{\beq}{\begin{equation}}
\newcommand{\eeq}{\end{equation}}
\newcommand{\bea}{\begin{eqnarray}}
\newcommand{\eea}{\end{eqnarray}}

\newcommand{\U}{{\rm U}}

\newcommand{\Sp}{{\rm Sp}}
\newcommand{\USp}{{\rm USp}}
\newcommand{\Spin}{\rm Spin}

\newcommand{\Z}{\mathbb Z}
\newcommand{\id}{\hbox{1\kern-.27em l}}
\newcommand{\tensor}[2]{\makebox[.5cm]{$#1$} \otimes \makebox[.5cm]{$#2$}}
\newcommand{\cE}{{\cal E}}
\newcommand{\cF}{{\cal F}}
\newcommand{\cR}{{\cal R}}
\newcommand{\D}{{\nabla}}

\long\def\symbolfootnote[#1]#2{\begingroup \def\thefootnote{\fnsymbol{footnote}}\footnote[#1]{#2}\endgroup} 

\begin{document}

\pagestyle{empty}

\begin{center}
\vspace*{30mm}
{\LARGE $(2,0)$ theory on circle fibrations}

\vspace*{20mm}
{\large
Hampus Linander\symbolfootnote[1]{hampus.linander@chalmers.se} and Fredrik Ohlsson\symbolfootnote[2]{fredrik.ohlsson@chalmers.se}
}

\vspace*{5mm}
Department of Fundamental Physics\\
Chalmers University of Technology\\
S-412 96 G\"oteborg, Sweden

\vspace*{30mm}{\bf Abstract:}
\end{center}
We consider $(2,0)$ theory on a manifold $M_6$ that is a fibration of a spatial $S^1$ over some five-dimensional base manifold $M_5$. Initially, we study the free $(2,0)$ tensor multiplet which can be described in terms of classical equations of motion in six dimensions. Given a metric on $M_6$ the low energy effective theory obtained through dimensional reduction on the circle is a Maxwell theory on $M_5$. The parameters describing the local geometry of the fibration are interpreted respectively as the metric on $M_5$, a non-dynamical $\U(1)$ gauge field and the coupling strength of the resulting low energy Maxwell theory. We derive the general form of the action of the Maxwell theory by integrating the reduced equations of motion, and consider the symmetries of this theory originating from the superconformal symmetry in six dimensions.  Subsequently, we consider a non-abelian generalization of the Maxwell theory on $M_5$. Completing the theory with Yukawa and $\phi^4$ terms, and suitably modifying the supersymmetry  transformations, we obtain a supersymmetric Yang-Mills theory which includes terms related to the geometry of the fibration.

\newpage
\pagestyle{plain}

\hrule
\tableofcontents
\vspace*{6mm}
\hrule

\section{Introduction}
The six-dimensional $(2,0)$ theory~\cite{Witten:1996, Witten:2004} has been a subject of much interest recently, in particular through its interpretation as the world volume theory of multiple $M5$-branes and the progress towards a better understanding of various brane configurations in $M$-theory. In~\cite{Seiberg:1997} the connection, through dimensional reduction on a circle, with supersymmetric Yang-Mills theory in five dimensions is discussed for the case of a six-dimensional manifold that is a direct product of $S^1$ and five-dimensional Minkowski space-time. Recent work exploring this connection between $(2,0)$ theory and supersymmetric Yang-Mills theory in five dimensions includes~\cite{Lambert:2010, Lambert:2011a, Douglas:2011,Lambert:2011b,Tachikawa:2011,Kim:2011,Singh:2011}. In the present paper we consider, along the lines of~\cite{Witten:2009}, the generalization of the above geometry to a general circle fibration $M_6 \to M_5$ for the free theory of the $(2,0)$ tensor multiplet~\cite{Strathdee:1987}. We also consider a non-abelian generalization of the Maxwell theory and construct a candidate for the interacting $(2,0)$ theory on $M_5$, which cannot be directly obtained by dimensional reduction.

The self-duality of the tensor three-form field strength makes a Lagrangian description problematic, but the tensor multiplet possesses a classical description in terms of equations of motion. The low energy effective theory obtained in the dimensional reduction on the $S^1$ fibre is a Maxwell theory on $M_5$ describing an abelian gauge field, five scalars and four spinors satisfying appropriate reality conditions. The coupling strength of the gauge theory, given by the square root of the radius of the $S^1$, is a function on the base manifold of the fibration. Furthermore, the $\U(1)$ subgroup of diffeomorphisms of $M_6$, corresponding to reparametrizations along the circle, gives rise to an additional non-dynamical abelian gauge field (the connection on the $\U(1)$ bundle $M_6$ over $M_5$) coupled to the gauge theory. In five dimensions it is possible to integrate the equations of motion to obtain an action describing the complete gauge theory, which we derive for a generic metric on $M_6$. In particular, the action contains terms including the $\U(1)$ field strength and the gradient of the radius in addition to the topological term for the gauge fields discussed in~\cite{Witten:2009}.

The theory of the $(2,0)$ tensor multiplet depends only on the conformal structure of $M_6$, i.e.~the equations of motion are covariant under a Weyl rescaling of the metric and simultaneous rescalings of the fields according to their conformal weight. A consequence of this conformal symmetry in six dimensions is that the gauge theory on $M_5$ obtained by the reduction is invariant under the corresponding simultaneous conformal rescalings of the five-dimensional metric, the dynamical fields and the (varying) coupling strength parameter.

When the manifold $M_6$ admits conformal Killing spinors the theory in six dimensions is also supersymmetric at the level of the equations of motion. (That is, the set of solutions to the equations of motion is closed under supersymmetry transformations.) In this case the same is true also for the five-dimensional theory. Furthermore, the action on $M_5$ is invariant under the same supersymmetry transformations as the equations of motion. In principle (if not in practice) this is a non-trivial feature since it extends the supersymmetry from its stationary points to the full action functional. Of course, the conformal symmetry of the tensor multiplet theory persists regardless of the existence of non-trivial conformal Killing spinors on $M_6$. We will, as previously mentioned, consider the reduction on arbitrary circle fibration, which implies that generically the six-dimensional theory will not be supersymmetric.

It is possible to generalize the abelian theory on $M_5$ to a non-abelian gauge theory by covariantizing the action and adding interaction terms. Including the ordinary Yukawa coupling and quartic scalar self-interaction of ordinary supersymmetric Yang-Mills theory produces a theory on $M_5$ that is invariant under the same generalized conformal symmetry as in the abelian case. Furthermore, when the ordinary non-linear term in the fermionic supersymmetry variation is included, the non-abelian gauge theory is supersymmetric whenever $M_5$ admits non-trivial solutions to the dimensionally reduced conformal Killing spinor equation, providing a non-trivial check of the construction.

The paper is organized as follows: In section~\ref{sec:Abelian(2,0)TheoryInSixDimensions} we consider the properties, in particular the superconformal symmetry, of the $(2,0)$ tensor multiplet theory in six dimensions. As is well known, the self-dual three form field does not admit a Lagrangian description (without the introduction of auxiliary fields \cite{Pasti:1997,Bandos:1997}) and we will thus only consider $H$ at the level of equations of motion. The scalar and spinor of the tensor multiplet, however, can be described using action functionals. In section~\ref{sec:SpatialCircleFibrations} we discuss the details of the spatial circle fibration and give explicit expressions for various geometrical quantities. We also describe how the Clifford algebra and spinors in six dimensions are decomposed with respect to the fibration. Subsequently, in section~\ref{sec:MaxwellTheoryInFiveDimensions} we perform the dimensional reduction and obtain Maxwell theory as the low energy effective theory in five dimensions. We also construct an action corresponding to the equations of motion of this theory and consider its properties under conformal rescalings and supersymmetry transformations in five dimensions. Finally, in section~\ref{sec:TheNonAbelianGeneralization}, we consider extending the supersymmetric Maxwell theory to a non-abelian supersymmetric Yang-Mills theory by adding interaction terms.

\section{Abelian $(2,0)$ theory in six dimensions}
\label{sec:Abelian(2,0)TheoryInSixDimensions}
In this section we review some aspects of the starting point for our consideration: The free tensor multiplet of $(2,0)$ theory on a six-dimensional manifold $M_6$ with local coordinates $y^M$, where $M=0,1\ldots,5$. The abelian $(2,0)$ theory depends only on the conformal structure on $M_6$, i.e.~on the conformal class of the metric $G_{MN}$ defined by the equivalence relation
\beq
\label{eqn:ConformalTransformaitionMetric}
G_{MN}(y) \sim e^{-2 \sigma(y)} G_{MN}(y)
\eeq
for some arbitrary function $\sigma(y)$, the coordinate dependence of which will be left implicit below. The invariance of the theory under $G_{MN} \to e^{-2\sigma} G_{MN}$ require suitable rescalings of the fields of the abelian multiplet which are discussed below. In addition to the conformal rescaling symmetry, there is a global R-symmetry group\footnote{Here, $\USp(4)$ is the compact real form of the Symplectic group $\Sp(4,\mathbb{C})$ with Lie algebra $C_2$.}
\beq
\cR \cong \USp(4) \cong \Spin(5)\,.
\eeq
Associated to the fundamental representation $\mathbf{4}$ of this group there is a symplectic metric $M_{\alpha \beta}$, where $\alpha$ and $\beta$ are spinor indices in the $\mathbf{4}$ representation. Details on this symplectic structure are provided in the appendix.

The manifold $M_6$ will generically have non-vanishing curvature and in order to describe spinor fields we will therefore need to introduce the vielbein $E_M^A$, defining locally an orthonormal frame. Here, $A=0,1,\ldots,5$ are flat indices, raised and lowered using the Minkowski metric $\eta_{AB}$, on which local Lorentz transformations act. The vielbein and its inverse $E^M_A$ satisfy the relations
\bea
\label{eqn:VielbeinProperties}
E_M^A E_{NA} = G_{MN} & , & E^M_A E_{MB} = \eta_{AB} \,,
\eea
which imply that under conformal transformations of the metric, the vielbein transforms as $E_M^A \to e^{-\sigma} E_M^A$.

\subsection{The tensor multiplet}
\label{sec:TheAbelianTensorMultiplet}
The field content of the abelian theory is a real three-form tensor field $H$, a scalar $\Phi^{\alpha \beta}$ and a positive chirality space-time spinor field $\Psi^{\alpha}$. The spinor and scalar fields transform respectively in the fundamental $\mathbf{4}$ representation and the $\mathbf{5}$ vector representation of the $\USp(4)$ R-symmetry. We will use the description of the fundamental representation as a spinor of $\Spin(5)$ and the vector representation as an antisymmetric bispinor which is traceless with respect to $M_{\alpha \beta}$. Further details regarding the symplectic transformation and reality properties of the fields are given in the appendix, where we also describe our conventions for Lorentz spinors in both $4+1$ and $5+1$ dimensions.

The tensor field transforms trivially under R-symmetry and is closed and self-dual, i.e.~it satisfies the equations of motion
\bea
\label{eqn:EquationOfMotionH}
dH = 0 & , & H = *_G H\,,
\eea
where the subscript $G$ on the Hodge dual indicates the six-dimensional metric. Both equations are conformally invariant by virtue of the metric independence of $dH=0$ and the determinant factor $\sqrt{-G}$ in the definition of the Hodge dual. The condition that $H$ be closed can be viewed as a consequence of it being the field strength of a two-form abelian gauge field $B$. However, for the purpose of the considerations of the present paper it is sufficient to consider only the three-form $H$, in which case $dH=0$ is considered as an equation of motion\footnote{Of course, $dH=0$ implies that $H=dB$ at least locally.}. As mentioned in the introduction there is no Lagrangian description of the self-dual tensor field in six dimensions.

The scalar $\Phi^{\alpha \beta}$ satisfies the symplectic reality condition $(\Phi^{\alpha \beta})^* = \Phi_{\alpha \beta}$ and the equation of motion 
\beq
\label{eqn:EquationOfMotionPhiSixDimensions}
G^{MN} \hat{\D}_M \hat{\D}_N \Phi^{\alpha \beta} + c\hat{R}\Phi^{\alpha \beta} = 0 \,,
\eeq
where $\hat{R}$ is the curvature scalar of the metric $G_{MN}$ and $\hat{\D}_M$ is the covariant derivative on $M_6$. In $d$ dimensions, this equation transforms covariantly under a conformal rescaling of the metric and a simultaneous transformation $\Phi^{\alpha \beta} \to e^{2 \sigma} \Phi^{\alpha \beta}$ of the scalar field, provided that 
\beq
c = -\frac{1}{4}\frac{(d-2)}{(d-1)} = -\frac{1}{5}\,,
\eeq
where in the last step we have inserted $d=6$. In contrast to the three-form, there is a Lagrangian description for the scalar; the equations of motion follow from the action
\beq
\label{eqn:ActionPhiSixDimensions}
S_{\Phi} = \int d^6y \sqrt{-G} \, \left( - \hat{\D}_M \Phi_{\alpha \beta} \hat{\D}^M \Phi^{\alpha \beta} + c\hat{R}\Phi_{\alpha \beta} \Phi^{\alpha \beta} \right) \,,
\eeq
which is real and invariant under Lorentz transformations, symplectic transformations and conformal rescalings.

Finally, as mentioned above, the fermionic degrees of freedom of the tensor multiplet are contained in a positive chirality spinor $\Psi^{\alpha}$ of the local Lorentz group. The spinor also satisfies a symplectic Majorana condition
\beq
\label{eqn:SymplecticMajoranaConditionSixDimensions}
(\Psi^\alpha )^* = M_{\alpha \beta} B_{(6)} \Psi^{\beta} \,,
\eeq
where $B_{(6)}$ is related to the charge conjugation matrix in six dimensions. The equation of motion for the spinor is the ordinary Dirac equation
\beq
\label{eqn:EquationOfMotionPsiSixDimensions}
\Gamma^M \hat{\D}_M \Psi^{\alpha} = 0 \,,
\eeq
where the curved space-time index $\Gamma$-matrices $\Gamma^M = \Gamma^A E_A^M$ are obtained using the vielbein as usual and the covariant derivative acting on the spinor $\Psi^{\alpha}$ is given by
\beq
\label{eqn:CovariantDerivativeSpinorSixDimensions}
\hat{\D}_M \Psi^{\alpha} = \partial_M \Psi^{\alpha} + \frac{1}{4} \Omega_M^{AB}\Gamma_{AB} \Psi^{\alpha} \,.
\eeq
Here, $\Omega_M^{A B}$ is the spin connection, i.e. the gauge field of local Lorentz transformations, of the six-dimensional manifold $M_6$. The Dirac equation is conformally covariant requiring that the spinor is rescaled as $\Psi^{\alpha} \to e^{\frac{5}{2} \sigma}\Psi^{\alpha}$, and may be obtained as the stationary point of the action
\beq
\label{eqn:ActionPsiSixDimensions}
S_{\Psi} = \int d^6y \sqrt{-G} \, i \overline{\Psi}_{\alpha} \Gamma^M \hat{\D}_M \Psi^{\alpha} \,.
\eeq
Using the reality properties of symplectic Majorana bilinears given in the appendix one verifies the hermiticity of this conformally invariant action functional, which is also a Lorentz and $\USp(4)$ scalar. Finally, we note that $\Psi^{\alpha}$ describes eight fermionic on-shell degrees of freedom, matching the number of bosonic ones (five for the scalar $\Phi^{\alpha \beta}$ and three for the tensor field $H$) as required for supersymmetry.

It should be emphasized that the absence of a Lagrangian formulation implies that our treatment of the tensor multiplet through its equations of motion is strictly classical. (It is possible to construct a Lagrangian for the self-dual tensor field through the introduction of an auxiliary scalar field~\cite{Pasti:1997, Bandos:1997}. However, a path integral quantization using such a Lagrangian appears to be problematic and we will refrain from considering the quantum theory in this paper.)

\subsection{Supersymmetry of the tensor multiplet}
\label{sec:SupersymmetryOfTheTensorMultiplet}
We now turn our attention to the supersymmetry variations of the $(2,0)$ tensor multiplet fields, given by
\beq
\label{eqn:SUSYVariationH}
\delta H_{MNP} = 3 \hat{\D}_{[M} \left( \overline{\Psi}_{\alpha} \Gamma_{NP]} \cE^{\alpha} \right) \,,
\eeq
\beq
\label{eqn:SUSYVariationPhiSixDimensions}
\delta \Phi^{\alpha \beta} = 2 \overline{\Psi}{}^{[\alpha} \cE^{\beta]} - \frac{1}{2}T^{\alpha \beta} \overline{\Psi}_{\gamma} \cE^{\gamma}
\eeq
and
\beq
\label{eqn:SUSYVariationPsiSixDimensions}
\delta \Psi^{\alpha} = \frac{i}{12} H_{MNP} \Gamma^{MNP}\cE^{\alpha} +  2 i M_{\beta \gamma} \hat{\D}_M \Phi^{\alpha \beta} \Gamma^M \cE^{\gamma} +  \frac{4i}{3} M_{\beta \gamma} \Phi^{\alpha \beta} \Gamma^M \hat{\D}_M \cE^{\gamma} \,,
\eeq
where the parameter $\cE^{\alpha}$ is a symplectic Majorana spinor of negative chirality in the $\mathbf{4}$ of $\USp(4)$. The variations $\delta H_{MNP}$, $\delta \Phi^{\alpha \beta}$ and $\delta \Psi^{\alpha}$  satisfy the same equations of motion as the original fields if (\ref{eqn:EquationOfMotionH}), (\ref{eqn:EquationOfMotionPhiSixDimensions}) and (\ref{eqn:EquationOfMotionPsiSixDimensions}) are imposed and the parameter $\cE^{\alpha}$ satisfies the conformal Killing spinor equation
\beq
\label{eqn:ConformalKillingSpinorEquationSixDimensions}
P_M \cE^{\alpha} = \hat{\D}_M \cE^{\alpha} - \frac{1}{d} \Gamma_M \Gamma^N \hat{\D}_N \cE^{\alpha} = 0 \,,
\eeq
which is conformally covariant in any dimension $d$ provided a rescaling of the parameter according to $\cE \to e^{-\frac{1}{2}\sigma} \cE$. The existence of non-vanishing $\cE$ satisfying the (\ref{eqn:ConformalKillingSpinorEquationSixDimensions}) imposes a non-trivial condition on the geometry of the manifold $M_6$. In order to have $(2,0)$ supersymmetry we must consequently restrict our attention to manifolds for which the kernel of the operator $P_M$ is non-trivial.
 
\section{Spatial circle fibrations}
\label{sec:SpatialCircleFibrations}
We now proceed to consider the case where $M_6$ is a fibration of $S^1$ over some five-dimensional base manifold $M_5$ of Lorentzian signature. Thus the curved vector index in six dimensions is split according to $M=(\mu,\varphi)$, where $\mu=0,\ldots,4$. We will allow ourselves to abuse the notation slightly by introducing $x^{\mu} = y^{\mu}$ and $\varphi = y^{\varphi}$. Here, $\varphi$ is the local coordinate along the $S^1$ fibre, while $x^{\mu}$ parametrize the base manifold $M_5$. We will adopt the convention that the range of the periodic $S^1$ coordinate is $0\leq\varphi<2\pi$. In the following section we will consider the dimensional reduction of the theory of the $(2,0)$ tensor multiplet on the $S^1$ to a (supersymmetric) Maxwell theory on $M_5$. The present section sets the stage for this reduction by investigating the various consequences of the specialization to a geometry in six dimensions that is a circle fibration.  

\subsection{Geometry of the fibration}
\label{sec:GeometryOfTheFibration}
The most general form of the metric on $M_6$ with the above decomposition is
\beq
\label{eqn:MetricCircleFibration}
ds^2 = g_{\mu \nu}dx^{\mu} dx^{\nu} + r^2 \left( d\varphi + \theta_{\mu}dx^{\mu} \right)^2\,.
\eeq
The fact that $M_6$ can be described as a $\U(1)$-bundle over $M_5$ implies the existence of an isometry along the $S^1$ and consequently the coefficient functions of (\ref{eqn:MetricCircleFibration}) are all independent of the coordinate $\varphi$. Thus, $g_{\mu \nu}(x)$ can be interpreted as the metric on $M_5$, $r(x)$ as the radius of the $S^1$ fibre and the vector $\theta_{\mu}(x)$ as an angular parameter. The special case when $\partial_{\mu} r = 0$ and $\theta_{\mu} = 0$ is referred to as the product metric. For generic $r(x)$ and $\theta_{\mu}(x)$ we can read of the component expressions for the decomposition of the metric
\beq
G_{\mu\nu}= g_{\mu\nu} + r^2\theta_{\mu}\theta_{\nu} \,\,\, , \,\,\, G_{\mu \varphi} = r^2\theta_{\mu} \,\,\, , \,\,\, G_{\varphi \varphi} = r^2
\eeq
and its inverse
\beq
G^{\mu\nu} =  g^{\mu\nu} \,\,\, , \,\,\, G^{\mu \varphi} = -\theta^{\mu} \,\,\, , \,\,\, G^{\varphi \varphi} = \frac{1}{r^2}+g^{\mu\nu}\theta_{\mu}\theta_{\nu} \,.
\eeq

In analogy with the curved index $M$, the flat vector index is split according to $A=(a,5)$ with $a=0,\ldots,4$. By a local Lorentz transformation the components of the vielbein $E_M^A$ and its inverse can be cast in the form
\beq
E^a_{\mu} = e^a_{\mu} \,\,\, , \,\,\, E_{\mu}^5 = r \theta_{\mu} \,\,\, , \,\,\, E_{\varphi}^a = 0 \,\,\, , \,\,\, E_{\varphi}^5 = r
\eeq
and
\beq
E_a^{\mu} = e_a^{\mu} \,\,\, , \,\,\, E^{\mu}_5 = 0 \,\,\, , \,\,\, E^{\varphi}_a = -\theta_{\mu}e^{\mu}_a \,\,\, , \,\,\, E^{\varphi}_5 = \frac{1}{r} \,,
\eeq
where $e_{\mu}^a$ is the vielbein on $M_5$, which satisfies the conditions (\ref{eqn:VielbeinProperties}). From their definition in terms of the vielbein it is now straightforward to compute the expressions for the non-vanishing components of the Levi-Civita connection $\hat{\Gamma}_{M N}^P$ and spin connection $\Omega_M^{A B}$ and obtain
\bea
\label{eqn:DimensionalReductionLeviCivitaConnection}
\hat{\Gamma}_{\mu \nu}^{\rho} & = & \Gamma_{\mu \nu}^{\rho} - r^2 {\cF^{\rho}}_{(\mu} \theta_{\nu)} - r \D^{\rho}r \theta_{\mu}\theta_{\nu} \cr
\hat{\Gamma}_{\mu \nu}^{\varphi} & = &\D_{(\mu} \theta_{\nu)} + r^2 \theta_{\rho} {\cF^{\rho}}_{(\mu} \theta_{\nu)} + r \theta_{\rho}\D^{\rho}r \theta_{\mu}\theta_{\nu} + 2\frac{1}{r}\D_{(\mu}r \theta_{\nu)} \cr
\hat{\Gamma}_{\mu \varphi}^{\rho} & = & \frac{1}{2} r^2 {\cF_{\mu}}^{\rho} - r \theta_{\mu}\D^{\rho}r \cr
\hat{\Gamma}_{\mu \varphi}^{\varphi} & = &  -\frac{1}{2} r^2 {\cF_{\mu}}^{\rho} \theta_{\rho} + r \theta_{\mu} \theta_{\rho}\D^{\rho}r + \frac{1}{r} \D_{\mu}r \cr
\hat{\Gamma}_{\varphi \varphi}^{\rho} & = & -r \D^{\rho}r \cr
\hat{\Gamma}_{\varphi \varphi}^{\varphi} & = & r \theta_{\rho} \D^{\rho}r
\eea
and
\bea
\label{eqn:DimensionalReductionSpinConnection}
\Omega_{\mu}^{ab} = \omega_{\mu}^{ab} - \frac{1}{2}r^2 \theta_{\mu}e_{\rho}^a e_{\sigma}^b \cF^{\rho \sigma} & , & \Omega_{\mu}^{a5} = \frac{1}{2}r e_{\nu}^a{\cF_{\mu}}^{\nu} - \theta_{\mu} e_{\nu}^a \D^{\nu}r \cr
\Omega_{\varphi}^{ab} = - \frac{1}{2}r^2 e_{\rho}^a e_{\sigma}^b \cF^{\rho \sigma} & , & \Omega_{\varphi}^{a5} = - e_{\nu}^a \D^{\nu}r \,,
\eea
where $\Gamma^{\rho}_{\mu \nu}$, $\omega_{\mu}^{a b}$ and $\D_{\mu}$ are respectively the Levi-Civita connection, the spin connection and the covariant derivative on the five-dimensional base manifold $M_5$. Finally, we can also compute an expression for the curvature scalar $\hat{R}$ appearing in the action (and equation of motion) for the scalar field in six dimensions. Using the expressions in (\ref{eqn:DimensionalReductionLeviCivitaConnection}) we obtain
\beq
\hat{R} = R - \frac{1}{4}r^2 \cF_{\mu\nu}\cF^{\mu\nu} - 2 \frac{1}{r} \D_{\mu}\D^{\mu}r \,,
\eeq
where $R$ denotes the curvature scalar of the metric $g_{\mu \nu}$ on $M_5$.

In the expressions above we have introduced the quantity
\beq
\label{eqn:FieldStrengthTheta}
\cF_{\mu \nu} = \partial_{\mu} \theta_{\nu} - \partial_{\nu} \theta_{\mu}\,,
\eeq
which from the point of view of the dimensionally reduced theory on $M_5$ can be interpreted as the field strength of the non-dynamical $\U(1)$ gauge field $\theta_{\mu}(x)$ corresponding to reparametrization invariance along the $S^1$ of the six-dimensional theory. Consequently, all physical five-dimensional quantities must be invariant under a $\U(1)$ gauge transformation $\theta_{\mu} \to \theta_{\mu} + \partial_{\mu}\lambda$, corresponding to coordinate transformation $\varphi \to \varphi + \lambda(x)$ in six dimensions, which generically implies that they can only depend on the gauge invariant field strength $\cF_{\mu \nu}$.

\subsection{Decomposition of spinors}
\label{sec:DecompositionOfSpinors}
The dimensional reduction of the $(2,0)$ theory will involve the decomposition of spinors and $\Gamma$-matrices in six dimensions in terms their five-dimensional counterparts. Since the dimension $2^{[d/2]}$ of a Dirac spinor is different in five and six dimensions this decomposition involves, in addition to the split of vector indices described above, a corresponding split of the Lorentz spinor index. We choose a representation of the six-dimensional Clifford algebra in terms of the tensor products involving the five-dimensional $\gamma$-matrices as
\beq
\label{eqn:GammaMatrixDecomposition}
\left\{
\begin{array}{ccc}
\Gamma^a & = & \tensor{\gamma^a}{\rho_1} \cr
\Gamma^5 & = & \tensor{\id}{\rho_2}
\end{array}
\right. \,,
\eeq
where $\rho_1$ and $\rho_2$ are the first two Pauli matrices. These satisfy $\rho_i^2 = \id$, $\rho_i^{\dagger} = \rho_i$ and $\{\rho_i,\rho_j\}=2\delta_{i j}$, for $i,j=1,2$, and consequently furnish a representation of the two-dimensional Euclidean Clifford algebra. We define the chirality operator in two Euclidean dimensions to be
\beq
\label{eqn:ChiralityOperatorTwoDimensions}
\rho = - i \rho_1 \rho_2 \,,
\eeq
where the overall sign is a matter of convention and the particular choice above will prove convenient in what follows. As a basis of two-dimensional spinors we may take the two eigenvectors $\eta_{\pm}$ of $\rho$, satisfying $\rho \eta_{\pm} = \pm \eta_{\pm}$, which can be chosen to be real and orthonormal. The action of the $\rho_i$ on these basis vectors is given by
\bea
\rho_1 \eta_{+} = \eta_{-} & , & \rho_2 \eta_{+} = i \eta_{-} \cr
\rho_1 \eta_{-} = \eta_{+} & , & \rho_2 \eta_{-} = - i \eta_{+} \,.
\eea
The charge conjugation matrix $C_{(6)}$ and the $B_{(6)}$ matrix are also decomposed as 
\beq
\label{eqn:ChargeConjugationMatrixDecomposition}
B_{(6)} = B_{(5)} \otimes \id \,\,\, , \,\,\, C_{(6)} = C_{(5)} \otimes \rho_1 \,, 
\eeq
which together with the decomposition (\ref{eqn:GammaMatrixDecomposition}) is consistent with the conventions for $\Gamma$-matrices in $5+1$ and $4+1$ dimensions. 

In the dimensional reduction from six to five dimensions the spinors decompose into tensor products in the same way as the $\Gamma$-matrices. Since the spinors $\Psi^{\alpha}$ and $\cE^{\alpha}$ relevant for the tensor multiplet and its supersymmetry are symplectic Majorana-Weyl, we will restrict considerations to a spinor $\Lambda^{\alpha}$ in the $\mathbf{4}$ representation of $\USp(4)$ which satisfies the symplectic Majorana condition (\ref{eqn:SymplecticMajoranaConditionSixDimensions}) and has a definite chirality $\Gamma \Lambda^{\alpha} = \pm \Lambda^{\alpha}$. With the conventions described in the appendix the chirality operator in six dimensions is $\Gamma=\id \otimes \rho$. Consequently, the decomposition of $\Lambda^{\alpha}$ is given by
\beq
\label{eqn:DimensionalReductionSpinor}
\Lambda^{\alpha} = \lambda^{\alpha} \otimes \eta_{\pm}
\eeq
according to its chirality. Note that we assume the symplectic spinor index $\alpha$ to be carried by the (Lorentz) spinor $\lambda^{\alpha}$, which is consistent with the fact that the R-symmetry is unchanged by the dimensional reduction, so that $\lambda^{\alpha}$ is also in the $\mathbf{4}$ of $\USp(4)$. Furthermore, it is consistent with the five-dimensional spinors satisfying the symplectic Majorana condition
\beq
\label{eqn:SymplecticMajoranaConditionFiveDimensions}
(\lambda^{\alpha})^* = M_{\alpha \beta} B_{(5)} \lambda^{\beta} \,,
\eeq
analogous to (\ref{eqn:SymplecticMajoranaConditionSixDimensions}), with the above decomposition of the charge conjugation matrix. Thus, (\ref{eqn:DimensionalReductionSpinor}) produces five-dimensional Lorentz spinors with the correct properties under symplectic transformations and complex conjugation. 

\section{Maxwell theory in five dimensions}
\label{sec:MaxwellTheoryInFiveDimensions}
We are now ready to consider the procedure at the heart of the present paper; the dimensional reduction on the $S^1$ fibre. In this section we consider the reduction of the theory of the free tensor multiplet. It is well known (see e.g.~\cite{Seiberg:1997}) that for a direct product of $S^1$ with five-dimensional Minkowski space, equipped with the product metric, this produces the ordinary maximally supersymmetric $N=4$ Maxwell theory in five dimensions at energies that are small compared to the fibre radius. In particular, the coupling is related to the (constant) radius of the $S^1$ fibre as $\tilde{g} = \sqrt{r}$. The R-symmetry of the Maxwell theory is the same as for the $(2,0)$ theory and the field content is a gauge field $A_{\mu}$ with field strength $F_{\mu \nu}$, a scalar $\phi^{\alpha \beta}$ and a symplectic Majorana spinor $\psi^{\alpha}$, the latter two transforming in the $\mathbf{5}$ and $\mathbf{4}$ representations of $\USp(4)$ respectively. The generalization to an arbitrary circle fibration should therefore in the low energy limit produce Maxwell theory with varying coupling strength, additional couplings to the non-dynamical $\U(1)$ gauge field $\theta_{\mu}$ and terms depending on the gradient of the radius $r(x)$. In the case when $M_6$ allows non-trivial solutions to $P_M \cE^{\alpha}=0$ there are unbroken supersymmetries of the $(2,0)$ theory and the five-dimensional theory should therefore be supersymmetric as well. 

Before deriving the complete action of the dimensionally reduced theory we review a consequence of the fact that $M_6$ is a fibration of $S^1$ over $M_5$ and the existence of local coordinates $(x^{\mu},\varphi)$ where $\varphi$ is a periodic coordinate along $S^1$. Collectively denoting the dynamical fields of the $(2,0)$ theory by $\Xi$, we can perform a Fourier expansion in $\varphi$
\beq
\Xi(x,\varphi) = \sum_{p \in \Z} \Xi_p(x) e^{i p \varphi} \,,
\eeq 
where $p$ is the momentum along $S^1$. The different Fourier modes constitute the Kaluza-Klein tower obtained in the reduction. All modes in the tower except the zero mode $\Xi_0$ acquire a mass, corresponding to the momentum along $S^1$. As we will see explicitly below, the curvature of $M_6$ will in fact introduce mass terms\footnote{The masses will be functions on $M_5$ rather than constants. However, they are uniquely determined by the conformal class of the metric on $M_6$.} also for the zero momentum Fourier modes. However, we will assume that the zero mode masses are negligible compared to the ones generated by non-zero momentum. Consequently, at sufficiently low energies in the reduced theory, the $p\neq0$ modes cannot be excited and therefore do not contribute to the low energy effective theory on $M_5$. The only remaining mode is thus the zero mode and the Fourier series is truncated $\Xi(x,\varphi) = \Xi_0(x)$. In particular, the fields are therefore independent of the fibre coordinate in the low energy limit that we are concerned with here. In what follows the dependence on the coordinates $x^{\mu}$ on $M_5$ is left implicit.

The condition $\partial_{\varphi} \Xi = 0$ is not covariant in six dimensions, which is not surprising since the Fourier expansion assumes explicitly the specific choice of local coordinates $y^M=(x^{\mu},\varphi)$. In order to obtain fields on $M_5$ that are suitably normalized it is also possible to rescale the Fourier modes with an arbitrary function of $x^{\mu}$. We will use this freedom below when we consider the reduction of the $(2,0)$ multiplet in the low energy limit.

\subsection{The Maxwell action on $M_5$}
\label{sec:TheMaxwellActionOnM5}
We consider first the scalar field $\Phi^{\alpha \beta}$ of the tensor multiplet. In this case there is an action in six dimensions which can be dimensionally reduced directly to produce the action in five dimensions. Using the freedom to introduce a relative scaling between the fields in five and six dimensions we let 
\beq
\label{eqn:DimensionalReductionPhi}
\Phi^{\alpha \beta} = \frac{1}{r \sqrt{2 \pi}} \phi^{\alpha \beta} \,,
\eeq
which implies that $\phi^{\alpha \beta}$ satisfies the same symplectic reality condition $(\phi^{\alpha \beta})^* = \phi_{\alpha \beta}$ as the six-dimensional scalar. Upon insertion in (\ref{eqn:ActionPhiSixDimensions}) and integration along the fibre coordinate (\ref{eqn:DimensionalReductionPhi}) yields
\beq
\label{eqn:ActionPhiFiveDimensions}
S_{\phi}  =  \int d^5x \sqrt{-g} \left( - \frac{1}{r} \D_{\mu}\phi_{\alpha \beta} \D^{\mu}\phi^{\alpha \beta} - \frac{1}{5} \frac{1}{r} R \phi_{\alpha \beta} \phi^{\alpha \beta} + K(g,r,\theta) \phi_{\alpha \beta} \phi^{\alpha \beta} \right)
\eeq
where we have introduced the quantity
\beq
K(g,r,\theta) = \frac{1}{r^3} \D_{\mu} r \D^{\mu} r  - \frac{3}{5} \frac{1}{r^2} \D_{\mu} \D^{\mu} r + \frac{1}{20} r \cF_{\mu\nu}\cF^{\mu\nu} \,,
\eeq
which contains information about the geometry, and in particular the curvature, of the manifold $M_6$.
The equation of motion for $\phi^{\alpha \beta}$ that follow from the action (\ref{eqn:ActionPhiFiveDimensions}) by construction agrees with the one obtained from dimensional reduction of the equations of motion (\ref{eqn:EquationOfMotionPhiSixDimensions}) in six dimensions as required.

Moving on to the spinors of the $(2,0)$ tensor multiplet we use the decomposition discussed in the previous section  to write
\beq
\label{eqn:DimensionalReductionPsi}
\Psi^{\alpha} = \frac{1}{r \sqrt{2 \pi}} \psi^{\alpha} \otimes \eta_{+} \,,
\eeq
which implies that $\psi^{\alpha}$ satisfies the symplectic Majorana reality condition
\beq
\label{eqn:SymplecticMajoranaConditionPsiFiveDimensions}
(\psi^{\alpha})^* = M_{\alpha \beta} B_{(5)} \psi^{\beta} \,.
\eeq
Once again we have introduced a rescaling to get canonically normalized spinors in five dimensions. The action (\ref{eqn:ActionPsiSixDimensions}) then yields
\beq
\label{eqn:ActionPsiFiveDimensions}
S_{\psi} = \int d^5x \sqrt{-g} \left( \frac{1}{r}i\overline{\psi}_{\alpha} \gamma^{\mu} \D_{\mu} \psi^{\alpha} - \frac{1}{8} \cF_{\mu \nu} \overline{\psi}_{\alpha} \gamma^{\mu\nu}\psi^{\alpha} \right)
\eeq
when integration over $S^1$ is performed, which entails the same equations of motion as obtained by dimensional reduction of the corresponding equations (\ref{eqn:EquationOfMotionPsiSixDimensions}) in six dimensions. 

In the case of the tensor $H_{M N P}$ the absence of an action implies that we must consider the equations of motion directly. The three-form $H$ can be decomposed as
\beq
H = E + F \wedge d\varphi = \frac{1}{3!}E_{\mu \nu \rho}dx^{\mu} \wedge dx^{\nu} \wedge dx^{\rho} + \frac{1}{2!}F_{\mu \nu}dx^{\mu} \wedge dx^{\nu} \wedge d\varphi \,.
\eeq
In the low energy limit we have $\partial_{\varphi} H_{MNP} = 0$ which in particular implies that the coefficients of $E_{\mu \nu \rho}$ and $F_{\mu \nu}$ are independent of $\varphi$, so that $E \in \Omega^3(M_5)$ and $F \in \Omega^2(M_5)$. Dimensional reduction of the equations of motion $dH=0$ and $H = *_G H$ in six dimensions then yields
\beq
dE = 0 \,\,\, , \,\,\, dF = 0
\eeq
and
\beq
\label{eqn:RelationEF}
E = - \frac{1}{r}*_{g}F + \theta \wedge F \,,
\eeq
where $*_g$ denotes the Hodge dual in five dimensions with respect to the metric $g$ and we have taken the liberty to denote the exterior derivative on $\Omega^*(M_5)$ by the same symbol as its six-dimensional counterpart. Using (\ref{eqn:RelationEF}) we can eliminate\footnote{This elimination is possible since the number of independent components are equal for a two-form $F$ and a three-form $E$ in five dimensions, and the same as the number of independent components of the self-dual three-form $H=*_G H$ in six dimensions.} $E$ from the theory on $M_5$, in which case $dE=0$ gives an equation of motion for $F$. The dimensional reduction of $H$ thus amounts to a two-form field strength $F$ on $M_5$ satisfying $dF=0$ and the equation of motion
\beq
d \left( \frac{1}{r}*_g F \right) - \cF \wedge F = 0\,.
\eeq
This equation of motion can, in contrast to that of $H$, be integrated to an action functional for the vector potential $A$ of which $F=dA$ is the field strength: 
\beq
\label{eqn:ActionFFiveDimensions}
S_{\mathrm{F}} = \int \left( - \frac{1}{r} F \wedge *_g F + \theta \wedge F \wedge F \right) \,.
\eeq
The complete Maxwell theory on $M_5$ obtained by dimensional reduction is thus described by the action
\beq
\label{eqn:CompleteActionFiveDimensions}
S = S_{\mathrm{F}} + S_{\psi} + S_{\phi} \,.
\eeq

Introducing $\tilde{g} = \sqrt{r(x)}$, in analogy with the case of a direct product manifold $M_6$, we see that (\ref{eqn:CompleteActionFiveDimensions}) describes a Maxwell theory with a coupling strength that is a function on $M_5$ as expected. The second part of the $S_{F}$ action is equivalent to the topological term given in equation (5.2) of~\cite{Witten:2009} in the sense that their variations are identical up to boundary terms. Furthermore, the complete action in five dimensions contains mass terms of geometrical origin, as mentioned in the beginning of this section. We see that requiring $R$, $\partial_{\mu}r$ and $\cF_{\mu \nu}$ to be sufficiently small ensures the consistency of the truncation of the Kaluza-Klein modes. Although the generic features of $S$ were previously known, the precise form of the action has to the best of our knowledge not been computed before.

\subsection{Conformal invariance}
\label{eqn:ConformalInvariance}
In conventional Maxwell theory with a constant coupling in five dimensions, the fact that $\tilde{g}$ is dimensionful implies that the theory is not conformally invariant. From the point of view of the reduction on $S^1$ the constant radius of the circle introduces a length scale that explicitly breaks the scale invariance of the six-dimensional theory. However, in the case of a general circle fibration we are currently considering, the coupling parameter is not restricted to be constant and consequently the conformal symmetry of the $(2,0)$ theory survives the reduction. From the decomposition (\ref{eqn:MetricCircleFibration}) of the metric we find that the geometric quantities scale according to
\beq
\label{eqn:ConformalRescalingGeometricalFiveDimensions}
g_{\mu \nu} \to e^{-2\sigma} g_{\mu \nu} \,\,\, , \,\,\, r \to e^{-\sigma} r \,\,\, , \,\,\, \theta_{\mu} \to \theta_{\mu}
\eeq
under a conformal transformation in six dimensions. Here, we restrict considerations to a parameter $\sigma$ that depends only on the coordinates on $M_5$ in order to obtain a conformal rescaling of the five-dimensional metric. This rescaling constitutes a generalized conformal symmetry of the Maxwell theory provided that the scalar and spinor fields are correspondingly rescaled according to (\ref{eqn:DimensionalReductionPhi}) and (\ref{eqn:DimensionalReductionPsi}) as
\beq
\label{eqn:ConformalRescalingFieldsFiveDimensions}
\phi^{\alpha \beta} \to e^{\sigma} \phi^{\alpha \beta} \,\,\, , \,\,\, \psi^{\alpha} \to e^{\frac{3}{2} \sigma} \psi^{\alpha} \,.
\eeq
From the point of view of the gauge theory on $M_5$ we must thus treat $r$ and $g_{\mu \nu}$ on equal footing, and consequently rescale not only the dynamical fields of the theory but also the coupling strength parameter $\tilde{g}$.

The Maxwell theory obtained in the reduction of the tensor multiplet is of course uniquely determined by the theory in six dimensions, but for the purpose of the considerations in the final section of this paper it is nevertheless interesting to consider the restrictions on an arbitrary gauge theory imposed by requiring the existence of generalized conformal invariance on $M_5$. In particular, given the canonically normalized kinetic term for the scalar $\phi^{\alpha \beta}$ it restricts the terms involving the gradient of the fibre radius and the five-dimensional curvature scalar, since these transform inhomogeneously under rescalings. (The inhomogeneous term produced by the kinetic term for the spinors $\psi^{\alpha}$ is proportional to $\overline{\psi}_{\alpha} \gamma^{\mu} \psi^{\alpha}$ which vanishes by symmetry.) However, terms involving $F_{\mu \nu}$ or $\cF_{\mu \nu}$ are invariant under conformal rescalings and therefore not restricted by this symmetry.

\subsection{Supersymmetry of the action}
\label{sec:SupersymmetryOfTheAction}
We can now restrict our attention to the case when the theory of the $(2,0)$ tensor multiplet in six dimensions is supersymmetric. As we saw above this amounts to requiring that the manifold $M_6$ admits non-trivial conformal Killing spinors $\cE^{\alpha}$ satisfying (\ref{eqn:ConformalKillingSpinorEquationSixDimensions}). Just as the dynamical fields of the Maxwell theory can be expanded in the periodic $\varphi$ coordinate we can expand $\cE^{\alpha}$ in a Fourier series as
\beq
\cE^{\alpha}(x^{\mu},\varphi) = \sum_{p \in \Z} \cE^{\alpha}_p(x) e^{i p \varphi} \,.
\eeq
There is however a significant difference: Being the parameter of supersymmetry transformations $\cE^{\alpha}$ is not a dynamical field and we can not simply integrate out the modes with non-zero momentum along $S^1$. However, acting on a dynamical field with a supersymmetry transformation involving any mode other than the zero mode $\cE^{\alpha}_0$ changes its mode number. In order to restrict considerations to supersymmetry transformations of the low energy effective theory we must therefore truncate the Fourier series of $\cE^{\alpha}$ and consider only the zero-mode $\cE_0^{\alpha}$. In this way we obtain the spinor parameter of supersymmetry of the low energy effective theory, which satisfies $\partial_{\varphi}\cE^{\alpha}=0$. Using the decomposition of spinors described in the previous section we then have
\beq
\cE^{\alpha} = \varepsilon^{\alpha} \otimes \eta_- \,,
\eeq
where $\partial_{\varphi} \varepsilon^{\alpha} = 0$ and according to (\ref{eqn:SymplecticMajoranaConditionFiveDimensions})
\beq
(\varepsilon^{\alpha})^* = M_{\alpha \beta} B_{(5)} \varepsilon^{\beta} \,.
\eeq
Dimensional reduction of the conformal Killing spinor equation (\ref{eqn:ConformalKillingSpinorEquationSixDimensions}) in addition yields the condition
\beq
\label{eqn:SUSYParameterConditionFiveDimensions}
\D_{\mu} \varepsilon^{\alpha} = \frac{1}{2} \frac{1}{r} \D^{\nu}r \gamma_{\mu}\gamma_{\nu} \varepsilon^{\alpha} + \frac{i}{8} r \cF^{\rho \sigma} \gamma_{\mu}\gamma_{\rho \sigma} \varepsilon^{\alpha} + \frac{i}{4} r {\cF_{\mu}}^{\nu}\gamma_{\nu} \varepsilon^{\alpha}
\eeq
on the five-dimensional spinor parameter $\varepsilon^{\alpha}$.

The supersymmetry transformation of the dynamical fields of the Maxwell theory on $M_5$ are obtained by dimensional reduction of the transformations (\ref{eqn:SUSYVariationH}), (\ref{eqn:SUSYVariationPhiSixDimensions}) and (\ref{eqn:SUSYVariationPsiSixDimensions}) in six dimensions, yielding
\beq
\label{eqn:SUSYVariationPhiFiveDimensions}
\delta \phi^{\alpha \beta} = 2 \overline{\psi}{}^{[\alpha} \varepsilon^{\beta]} - \frac{1}{2}T^{\alpha \beta} \bar{\psi}_{\gamma}\varepsilon^{\gamma} \,,
\eeq
\bea
\label{eqn:SUSYVariationFFiveDimensions}
\delta F_{\mu \nu} & = & - 2 i \D_{[\mu} \overline{\psi}_{\alpha} \gamma_{\nu]} \varepsilon^{\alpha} + i \frac{1}{r} \D^{\rho}r \overline{\psi}_{\alpha} \gamma_{\mu \nu \rho} \varepsilon^{\alpha} - 2 i \frac{1}{r} \D_{[\mu}r \overline{\psi}_{\alpha} \gamma_{\nu]} \varepsilon^{\alpha} \cr
&& + r \cF_{\mu \nu} \overline{\psi}_{\alpha} \varepsilon^{\alpha} + \frac{3}{2} r {\cF_{[\mu}}^{\rho} \overline{\psi}_{\alpha} \gamma_{\nu] \rho} \varepsilon^{\alpha} - \frac{1}{4} r \cF^{\rho \sigma}\overline{\psi}_{\alpha} \gamma_{\mu \nu \rho \sigma} \varepsilon^{\alpha}
\eea
and
\bea
\label{eqn:SUSYVariationPsiFiveDimensions}
\delta \psi^{\alpha} & = & \frac{1}{2} F_{\mu \nu} \gamma^{\mu \nu} \varepsilon^{\alpha} + 2 i M_{\beta \gamma} \D_{\mu} \phi^{\alpha \beta} \gamma^{\mu} \varepsilon^{\gamma} \cr
&& + \, 2i \frac{1}{r} M_{\beta \gamma} \phi^{\alpha \beta} \D_{\mu} r \gamma^{\mu} \varepsilon^{\gamma} -  r M_{\beta \gamma} \phi^{\alpha \beta} \cF_{\mu \nu} \gamma^{\mu \nu} \varepsilon^{\gamma} \,.
\eea
We note that $\cF$ and $\D_{\mu}r$ enter in the supersymmetry variation of $F_{\mu \nu}$ and $\psi^{\alpha}$ through the covariant derivative of $\varepsilon^{\alpha}$ and the relation (\ref{eqn:SUSYParameterConditionFiveDimensions}). It is a straightforward but somewhat laborious task to verify that the complete action (\ref{eqn:CompleteActionFiveDimensions}) is invariant under the transformations (\ref{eqn:SUSYVariationPhiFiveDimensions}), (\ref{eqn:SUSYVariationFFiveDimensions}) and (\ref{eqn:SUSYVariationPsiFiveDimensions}) provided that the supersymmetry parameter $\varepsilon^{\alpha}$ satisfies (\ref{eqn:SUSYParameterConditionFiveDimensions}). (Supersymmetry at the level of the equations of motion in five dimensions is an immediate consequence of supersymmetry in six dimensions.) In analogy to the case for the $(2,0)$ tensor multiplet, supersymmetry of the action thus imposes a non-trivial geometrical condition on the manifold $M_5$, namely the existence of non-trivial solutions to (\ref{eqn:SUSYParameterConditionFiveDimensions}).

\subsection{The product metric}
\label{sec:TheProductMetric}
In order to verify that the results derived in the present section reproduces the known result for $M_6 = M_5 \times S^1$ with the product metric, we will now consider the case $\theta_{\mu}(x) = 0$ and $\partial_{\mu} r(x) = 0$. In this case we expect to recover ordinary Maxwell theory on $M_5$, which we still allow to be arbitrary. From (\ref{eqn:ActionPhiFiveDimensions}), (\ref{eqn:ActionPsiFiveDimensions}) and (\ref{eqn:ActionFFiveDimensions}) we find that the action for the case of the product metric reduces to
\beq
\label{eqn:ActionTrivialFibration}
S = \frac{1}{\tilde{g}^2}\int d^5x \sqrt{-g} \left( - \frac{1}{2} F_{\mu \nu} F^{\mu \nu} + i\overline{\psi}_{\alpha} \gamma^{\mu} \D_{\mu} \psi^{\alpha} - \D_{\mu}\phi_{\alpha \beta} \D^{\mu}\phi^{\alpha \beta}  + c R \phi_{\alpha \beta} \phi^{\alpha \beta} \right)
\eeq
where we have identified the Maxwell coupling constant as $\tilde{g} = \sqrt{r}$, which has the appropriate dimension in five dimensions (and in this case is a proper constant). The equations of motion obtained from this action are
\beq
\label{eqn:EquationsOfMotionDirectProduct}
\D_{\mu}F^{\mu \nu} = 0 \,\,\, , \,\,\, \gamma^{\mu}\D_{\mu}\psi^{\alpha} = 0 \,\,\, , \,\,\, \D_{\mu}\D^{\mu}\phi^{\alpha \beta} + c R\phi^{\alpha \beta} = 0 \,.
\eeq
The appearance of terms proportional to $R$ is a consequence of the conformal symmetry in six dimensions. Note, however, that as discussed above, the radius $r$ explicitly breaks the scale invariance of the theory on $M_5$ as long as we consider it to be constant, since this condition eliminates the possibility of rescaling the coupling to compensate for the inhomogeneous transformation of $S$ under simultaneous rescalings (\ref{eqn:ConformalRescalingGeometricalFiveDimensions}) and (\ref{eqn:ConformalRescalingFieldsFiveDimensions}).

As a next step we consider requiring the existence of supersymmetry in the $(2,0)$ theory with product metric on $M_6 \to M_5$. The expressions (\ref{eqn:SUSYVariationPhiFiveDimensions}), (\ref{eqn:SUSYVariationFFiveDimensions}) and (\ref{eqn:SUSYVariationPsiFiveDimensions}) reduce to the familiar variations of supersymmetric Maxwell theory. Furthermore, the supersymmetry parameter must satisfy (\ref{eqn:SUSYParameterConditionFiveDimensions}), which for the product metric reduces to
\beq
\label{eqn:SUSParamterConditionDirectProduct}
\D_{\mu} \varepsilon^{\alpha} = 0 \,.
\eeq
Taking another covariant derivative and antisymmetrizing one obtains $\D_{[\mu}\D_{\nu]} \varepsilon^{\alpha} = 0$ which implies $R = 0$, so that the action (and the corresponding equations of motion) reduces to that of ordinary supersymmetric Maxwell theory on $M_5$.

\section{The non-abelian generalization}
\label{sec:TheNonAbelianGeneralization}
In the previous sections of this paper we have considered exclusively the free tensor multiplet of $(2,0)$ theory and the low energy Maxwell theory obtained by its reduction on the $S^1$ fibre of $M_6$. We would now like to extend our scope to consider also the $A_r$,$D_r$ and $E_r$ series of $(2,0)$~\cite{Witten:1996} in the circle fibration geometry. However, a direct derivation of the low energy theory on $M_5$ by dimensional reduction is not possible in this case because, unlike the free tensor multiplet, the ADE type $(2,0)$ theories have no classical field theory description in terms of equations of motion\footnote{In~\cite{Witten:2009} this is explained in terms of the absence of a classical notion of a gerbe with non-abelian structure group of which $H$ is the curvature.}.

However, we have some information regarding the low energy theory on $M_5$ obtained by reduction of the $(2,0)$ theory associated to a simply laced group $G$. Conformal invariance in six dimensions entails generalized conformal invariance, discussed above for the free tensor multiplet, and if $M_6$ admits non-trivial conformal Killing spinors parametrizing supersymmetry transformations the theory on $M_5$ will also be supersymmetric. Furthermore, the theory on $M_5$ should be a theory of gauge fields with gauge group $G$. In particular, for the case $M_6 = M_5 \times S^1$ with a product metric (i.e. $\theta_{\mu} = 0$ and $\partial_{\mu} r = 0$) the theory on $M_5$ is supersymmetric Yang-Mills theory with gauge group $G$\footnote{In the special case where $M_5$ is Minkowski the Yang-Mills theory is maximally supersymmetric.}. For a generic metric on $M_6$ it should be coupled to the background $\U(1)$ gauge field on $M_5$, corresponding to reparametrization invariance of the fibre.

The generalization of the theory described by the action (\ref{eqn:CompleteActionFiveDimensions}) thus involves promoting $A$ to the connection of a principal $G$-bundle over $M_5$ and $\phi^{\alpha \beta}$ and $\psi^{\alpha}$ to sections of associated adjoint bundles. The dynamical fields of the theory are consequently $A_{\mu}^a$, $\phi^{\alpha \beta}_a$ and $\psi^{\alpha}_a$ where we denote by $a$ the index in the adjoint representation of the Lie algebra $\mathfrak{g}$ of $G$. (Since we will not use local Lorentz vector indices on $M_5$ explicitly in this section this notational overlap will hopefully not cause any confusion.) With anti-hermitian Lie algebra generators we have the standard expressions for the gauge field strength and the covariant derivative of a field $\chi^a$ in the adjoint representation, given by
\beq
\label{eqn:GaugeFieldStrengthNonAbelian}
F_{\mu \nu}^a = \D_{\mu} A_{\nu}^a - \D_{\nu} A_{\mu}^a + {f^a}_{b c} A^b_{\mu} A^c_{\nu}
\eeq
and
\beq
\label{eqn:CovariantDerivativeNonAbelian}
D_{\mu} \chi^a = \D_{\mu} \chi^a + {f^a}_{b c} A^b_{\mu} \chi^c \,,
\eeq
where $f^{a b c}$ are the structure constants of $\mathfrak{g}$. As in the previous sections, the derivative $\D_{\mu}$ is covariant w.r.t.~both general coordinate transformations and local Lorentz transformations. We can then make the action (\ref{eqn:CompleteActionFiveDimensions}) gauge invariant by letting all fields transform in the adjoint representation of $\mathfrak{g}$, replacing the field strength and derivatives with gauge covariant ones and taking the trace in the adjoint representation, giving
\beq
\label{eqn:ActionPhiFiveDimensionsNonAbelian}
S_{\phi}  =  \int d^5x \sqrt{-g} \left( - \frac{1}{r} D_{\mu}\phi^a_{\alpha \beta} D^{\mu}\phi_a^{\alpha \beta} - \frac{1}{5} \frac{1}{r} R \phi^a_{\alpha \beta} \phi_a^{\alpha \beta} + K(g,r,\theta) \phi^a_{\alpha \beta} \phi_a^{\alpha \beta} \right) \,,
\eeq
\beq
\label{eqn:ActionPsiFiveDimensionsNonAbelian}
S_{\psi} = \int d^5x \sqrt{-g} \left( \frac{1}{r}i\overline{\psi}{}^a_{\alpha} \gamma^{\mu} D_{\mu} \psi_a^{\alpha} - \frac{1}{8} \cF_{\mu \nu} \overline{\psi}{}^a_{\alpha} \gamma^{\mu\nu}\psi_a^{\alpha} \right)
\eeq
and
\beq
\label{eqn:ActionFFiveDimensionsNonAbelian}
S_{\mathrm{F}} = \int \mathrm{tr} \left( - \frac{1}{r} F \wedge *_g F + \theta \wedge F \wedge F \right) \,.
\eeq
In the same way we also obtain gauge covariant supersymmetry variations
\beq
\label{eqn:SUSYVariationPhiFiveDimensionsNonAbelian}
\delta \phi_a^{\alpha \beta} = 2 \overline{\psi}{}_a^{[\alpha} \varepsilon^{\beta]} - \frac{1}{2}T^{\alpha \beta} \bar{\psi}_{\gamma a}\varepsilon^{\gamma} \,,
\eeq
\bea
\label{eqn:SUSYVariationFFiveDimensionsNonAbelian}
\delta F^a_{\mu \nu} & = & - 2 i D_{[\mu} \overline{\psi}{}^a_{\alpha} \gamma_{\nu]} \varepsilon^{\alpha} + i \frac{1}{r} D^{\rho}r \overline{\psi}{}^a_{\alpha} \gamma_{\mu \nu \rho} \varepsilon^{\alpha} - 2 i \frac{1}{r} D_{[\mu}r \overline{\psi}{}^a_{\alpha} \gamma_{\nu]} \varepsilon^{\alpha} \cr
&& + r \cF_{\mu \nu} \overline{\psi}{}^a_{\alpha} \varepsilon^{\alpha} + \frac{3}{2} r {\cF_{[\mu}}^{\rho} \overline{\psi}{}^a_{\alpha} \gamma_{\nu] \rho} \varepsilon^{\alpha} - \frac{1}{4} r \cF^{\rho \sigma}\overline{\psi}{}^a_{\alpha} \gamma_{\mu \nu \rho \sigma} \varepsilon^{\alpha}
\eea
and
\bea
\label{eqn:SUSYVariationPsiFiveDimensionsNonAbelian}
\delta \psi_a^{\alpha} & = & \frac{1}{2} F_{a \mu \nu} \gamma^{\mu \nu} \varepsilon^{\alpha} + 2 i M_{\beta \gamma} D_{\mu} \phi_a^{\alpha \beta} \gamma^{\mu} \varepsilon^{\gamma} \cr
&& + \, 2i \frac{1}{r} M_{\beta \gamma} \phi_a^{\alpha \beta} D_{\mu} r \gamma^{\mu} \varepsilon^{\gamma} -  r M_{\beta \gamma} \phi_a^{\alpha \beta} \cF_{\mu \nu} \gamma^{\mu \nu} \varepsilon^{\gamma} \,.
\eea
The condition (\ref{eqn:SUSYParameterConditionFiveDimensions}) receives no modification since the conformal Killing spinor equation on $M_6$ is satisfied by the supersymmetry parameter also for non-abelian $(2,0)$ theory. This is consistent, since $\varepsilon^{\alpha}$ and the parameters $r$ and $\theta_{\mu}$ are all invariant under $G$ gauge transformations.

In order to recover ordinary supersymmetric Yang-Mills theory in the case where $M_6 = M_5 \times S^1$ with product metric we must add a Yukawa term and a $\phi^4$ term to the action to obtain
\bea
\label{eqn:ActionNonAbelianExtension}
S_{\mathrm{YM}} & = & \ldots + \int d^5x \sqrt{-g} \left( 2 \frac{1}{r}  f^{a b c} M_{\alpha \gamma} M_{\beta \delta} \phi^{\alpha \beta}_a \overline{\psi}{}^{\gamma}_b \psi^{\delta}_c \right. \cr
&& \left. + \frac{1}{r} {f^{a b}}_e f^{c d e} M_{\sigma \alpha} M_{\beta \gamma} M_{\delta \lambda} M_{\tau \rho} \phi_a^{\alpha \beta} \phi_b^{\gamma \delta} \phi_c^{\lambda \tau} \phi_d^{\rho \sigma} \right) \,,
\eea
and modify the supersymmetry variation of the fermionic field with a non-linear term according to
\beq
\label{eqn:SUSYVariationNonAbelianExtension}
(\delta \psi^{\alpha}_a)_{\mathrm{YM}} = \ldots  + 2 {f_a}^{b c} M_{\beta \gamma} M_{\delta \lambda} \phi_b^{\alpha \beta} \phi_c^{\gamma \delta} \varepsilon^{\lambda} \,.
\eeq
The above action and supersymmetry variations transform correctly under generalized conformal rescalings and satisfy the appropriate reality conditions. By a straightforward computation (involving some rather lengthy R-symmetry manipulations) one verifies that when $M_5$ admits non-trivial solutions to (\ref{eqn:SUSYParameterConditionFiveDimensions}) the action (\ref{eqn:ActionNonAbelianExtension}) is invariant under the modified supersymmetry transformations. Thus, the model described by (\ref{eqn:ActionNonAbelianExtension}) and (\ref{eqn:SUSYVariationNonAbelianExtension}) constitutes a generalization of the Maxwell theory, obtained for the free tensor multiplet in the case of a general fibration of $S^1$ over $M_5$, to a non-abelian Yang-Mills theory with varying coupling strength, coupled to a background $\U(1)$ gauge field.

Just as in the case of the Maxwell theory, the non-vanishing right hand side of (\ref{eqn:SUSYParameterConditionFiveDimensions}), which from a five-dimensional point of view depends on the gradient of the coupling strength and the non-dynamical background gauge field, implies that the presence of the terms in (\ref{eqn:SUSYVariationNonAbelianExtension}) that depend on $\theta_{\mu}$ and $\partial_{\mu}r$ is required for supersymmetry of the Yang-Mills theory. For $\phi^{\alpha \beta}$ and $\psi^{\alpha}$, the terms in the action depending on $\partial_{\mu} r$ and $\cF_{\mu \nu}$, required for invariance under generalized conformal rescalings, are quadratic and consequently introduce no novel interactions. (The topological $\theta$-term, however, is quadratic in the non-linear field strength (\ref{eqn:GaugeFieldStrengthNonAbelian}) and does represent an interaction related to the fibration geometry.) In this sense, the model constitutes the minimal non-abelian extension of the Maxwell theory obtained for the tensor multiplet.

Since we have no field theory description of $(2,0)$ theory of type ADE it is not possible to verify that (\ref{eqn:ActionNonAbelianExtension}) and (\ref{eqn:SUSYVariationNonAbelianExtension}) gives the correct theory on $M_5$ by explicit computation of the reduction. However, it appears to be difficult to construct other non-abelian gauge theories with all the required properties due to the strong restrictions imposed by generalized conformal symmetry and supersymmetry on $M_5$.

\section{Summary and conclusion}
\label{sec:SummaryAndConclusion}
In this paper we first considered the dimensional reduction of the theory of a free $(2,0)$ tensor multiplet on a circle fibration $M_6 \to M_5$ in detail. The low energy effective theory obtained on $M_5$ is a Maxwell theory describing an abelian gauge field $A_{\mu}$ with field strength $F_{\mu \nu}$, a scalar $\phi^{\alpha \beta}$ and a spinor $\psi^{\alpha}$, where the latter fields transform respectively in the $\mathbf{5}$ and $\mathbf{4}$ of the $\USp(4)$ R-symmetry. For a generic metric on $M_6$ the coupling strength of the Maxwell theory is a function on $M_5$ given by the square root of the fibre radius $r(x)$. Furthermore, the Lagrangian contains quadratic terms for the scalar and spinor fields and a topological $\theta$-term for the gauge field, related to the local geometry of the fibration $M_6$. (In a path integral quantization of the gauge theory on $M_5$, the overall normalization of the action is determined by the requirement that the factor in the integrand containing the exponentiation of the topological $\theta$-term be well defined.) The terms are explicitly derived and the result given in (\ref{eqn:ActionPhiFiveDimensions}), (\ref{eqn:ActionPsiFiveDimensions}) and (\ref{eqn:ActionFFiveDimensions}).

The equations of motion of the full theory on $M_5$ can (in contrast to those of the $(2,0)$ theory on $M_6$) be integrated to an action functional. The action is invariant under generalized conformal rescalings of the metric, dynamical fields and the coupling strength. Furthermore, it is invariant under the supersymmetry transformations (\ref{eqn:SUSYVariationPhiFiveDimensions}), (\ref{eqn:SUSYVariationFFiveDimensions}) and (\ref{eqn:SUSYVariationPsiFiveDimensions}), obtained by reduction of the corresponding variations in six dimensions, when $M_5$ admits non-trivial solutions to (\ref{eqn:SUSYParameterConditionFiveDimensions}).

We also considered a non-abelian generalization of the Maxwell gauge theory in order to find the description of the dimensional reduction of ADE type $(2,0)$ theory on the $S^1$ fibre of $M_6$. We find that gauge covariantizing the abelian theory and including Yukawa and $\phi^4$ interaction terms produces a theory with the required invariance under generalized conformal rescalings on $M_5$. As a further consistency check we find that with a quadratic modification of the fermionic supersymmetry transformations the theory is supersymmetric if $M_5$ admits solutions to (\ref{eqn:SUSYParameterConditionFiveDimensions}). Finally, in the case of a product metric on $M_6$, corresponding to $\theta_{\mu} = 0$ and $\partial_ {\mu}r=0$ for the coupling strength and background gauge field in the five dimensional perspective, the generalization reduces to ordinary supersymmetric Yang-Mills theory. We emphasize that the gauge theory on $M_5$ is not directly derived from $(2,0)$ theory on $M_6$ but constitutes the minimal (in the sense described above) candidate for its reduction on $S^1$.

As discussed above, supersymmetry of the $(2,0)$ theory requires the existence of conformal Killing spinors, i.e. non-trivial solutions to (\ref{eqn:ConformalKillingSpinorEquationSixDimensions}) which in the special case of a circle fibration reduces to (\ref{eqn:SUSYParameterConditionFiveDimensions}) on the base $M_5$. The classification of manifolds of Lorentzian signature admitting conformal Killing spinors has been extensively studied (see e.g.~\cite{Baum:2002} and references therein). It would be interesting to investigate which of these classes contain circle fibrations.

An interesting example of a manifold that does admit conformal Killing spinors (in particular covariantly constant spinors) is discussed in~\cite{Witten:2009}: Let $M_6 = \mathbb{R}^{1,1} \times TN$, where $TN$ is the Taub-NUT hyper-K\"ahler space which admits a $\U(1)$ action that preserves its hyper-K\"ahler structure. However, the $\U(1)$ action has a fix-point at the origin of the $\mathbb{R}^3$ underlying the $TN$ and consequently the description of $M_6$ as a $\U(1)$-fibration becomes singular on $W = \mathbb{R}^{1,1} \times \{0\}$. Over $M_5 \backslash W$ the description of $M_6$ as a $\U(1)$ bundle is valid and the results of the present paper are applicable, but on $W$ the curvature $\cF$ has a singularity. In particular, this implies that the topological term, which can equivalently be expressed in terms of the Chern-Simons form, transforms anomalously under gauge transformations requiring the introduction of a WZW model localized on $W$ to cancel the anomaly. A natural extension of the present work would be to consider manifolds $M_6$ with codimension 4 singularities as in the example above and investigate the coupling of the WZW model to the gauge theory on $M_5$. We intend to pursue this direction in future work.

\vspace*{5mm}
During the final preparation of this manuscript~\cite{Gustavsson:2011} appeared, which treats in detail the case of a single $M5$ brane on $M_6 = \mathbb{R}^{1,2} \times S^3$ and the reduction on the Hopf fibration. Related results concerning instantons in the five-dimensional gauge theory are presented in~\cite{Lambert:2011c}.

The authors gratefully acknowledge M{\aa}ns Henningson for suggesting the problem and for many illuminating discussions and valuable advice. We have also benefitted from discussions with Bengt E.W. Nilsson, Ulf Gran and Martin Cederwall. This research was supported by grants from the Swedish Research Council and the G\"oran Gustafsson Foundation.

\newpage
\begin{appendix}
\section{Conventions}
\subsection{Symplectic transformation properties}
We first consider the symplectic transformation properties of the scalar and spinor fields, which fall in non-trivial representations of the $\cR$-symmetry group $\USp(4)$. We let $\alpha =1,2,3,4$ be the spinor index of the fundamental $\mathbf{4}$ representation of $\USp(4)$ and denote the symplectic structure by $M_{\alpha \beta}$ (we refrain from using the conventional notation $\Omega$ for the symplectic structure to avoid confusion with the spin connection).

We further denote by $V_\mathbf{4}$ and $V_\mathbf{\bar{4}}$ the dual modules of the $\mathbf{4}$ representation and its conjugate representation $\mathbf{\bar{4}}$, and let the vertical position of the index indicate the representation according to $v^{\alpha} \in V_{\mathbf{4}}$ and $w_{\alpha} \in V_{\mathbf{\bar{4}}}$. The fundamental representation and its conjugate are related under complex conjugation so we can infer that
\bea
(v^a)^* \in V_{\mathbf{\bar{4}}} & , & (w_a)^* \in V_{\mathbf{4}} \,.
\eea
While the two representations $\mathbf{4}$ and $\mathbf{\bar{4}}$ are unitarily equivalent, it is convenient to distinguish between them when considering reality conditions for the various fields of the $(2,0)$ theory. We will therefore distinguish upper and lower indices and utilize the fact that complex conjugation interchanges these two types of indices according to the relation above. 

The symplectic form is non-degenerate and antisymmetric, $M_{\alpha \beta} = -M_{\beta \alpha}$, and thus constitutes a metric on $V_{\mathbf{4}}$ providing an isomorphism $M : V_\mathbf{4} \to V_{\mathbf{\bar{4}}} $ between the vector space $V_{\mathbf{4}}$ and its dual. Similarly, its inverse $T^{\alpha \beta}$ defines an isomorphism $T : V_\mathbf{\bar{4}} \to V_{\mathbf{4}}$. The isomorphisms are given by
\bea
v_{\alpha} = M_{\alpha \beta} v^{\beta} & , & w^{\alpha} = T^{\alpha \beta} w_{\beta} \,,
\eea 
and because $T = M^{-1}$ they satisfy the relations
\bea
T^{\alpha \beta} M_{\beta \gamma} = {\delta^{\alpha}}_{ \gamma}  & , & M_{\alpha \beta} T^{\beta \gamma} = {{\bar{\delta}}^{\,\,\,\, \gamma}}_{\alpha} \,,
\eea
where ${\delta^{\alpha}}_{ \beta}$ and ${\bar{\delta}^{\,\,\,\, \beta}}_{\alpha}$ are the identity operators on $V_{\mathbf{4}}$ and $V_{\mathbf{\bar{4}}}$ respectively. Using the metric and its inverse we can thus raise and lower $\USp(4)$ spinor indices. Finally, the complex conjugate of the symplectic metric is given by $(M_{\alpha \beta})^* = M^{\alpha \beta} = - T^{\alpha \beta}$ and similarly for $T^{\alpha \beta}$.

The spinor field of the tensor multiplet transforms in the fundamental representation of $\USp(4)$ and consequently has a single $\USp(4)$ spinor index $\Psi^{\alpha}$. The scalar field of the multiplet, on the other hand, transforms in the vector representation $\mathbf{5}$, which is obtained from the antisymmetric part of the tensor product
\beq
\mathbf{4} \otimes \mathbf{4} = \mathbf{1} \oplus \mathbf{5} \oplus \mathbf{10}
\eeq
by imposing the vanishing of the antisymmetric trace constituting the singlet. The scalar thus is an antisymmetric bispinor $\Phi^{\alpha \beta} = - \Phi^{\beta \alpha}$ satisfying the tracelessness condition $M_{\alpha \beta} \Phi^{\alpha \beta} = 0$. Furthermore, the properties of $M_{\alpha \beta}$ allows us to impose a consistent symplectic reality condition on $\Phi^{\alpha \beta}$, given by
\beq
(\Phi^{\alpha \beta})^* = \Phi_{\alpha \beta} = M_{\alpha \gamma} M_{\beta \delta} \Phi^{\gamma \delta}\,.
\eeq
We also note that this condition is consistent with complex conjugation relating the $\mathbf{4}$ and $\mathbf{\bar{4}}$ representations. In the next subsection we will consider a symplectic reality condition for the spinor field $\Psi^{\alpha}$ as well. 

\subsection{Spinors in $4+1$ and $5+1$ dimensions}
Next, we consider the spinor representations of the Lorentz group in the dimensions relevant for the considerations of the present paper. We work in Lorentzian signature and use conventions where the flat Minkowski metric is $\eta = {\rm diag}(-1,1,\ldots,1)$. The Clifford algebra in $5+1$ and $4+1$ dimensions is  $\{\gamma^a,\gamma^b\} = 2 \eta^{ab}$ and $\{\Gamma^A,\Gamma^B\} = 2 \eta^{AB}$ respectively\footnote{Here, as before, $a$ and $A$ are flat vector indices in five and six dimensions respectively.}. The hermiticity properties of the $\Gamma$-matrices are given by
\bea
(\gamma^a)^{\dagger} & = & \gamma^0 \gamma^a \gamma^0 \cr
(\Gamma^A)^{\dagger} & = & \Gamma^0 \Gamma^A \Gamma^0 \,.
\eea
The charge conjugation matrix in the respective dimensions is uniquely determined (up to a complex phase) by the relations
\bea
C_{(5)}^T = - C_{(5)} & , & (\gamma^{a})^T = C_{(5)} \gamma^{a} C_{(5)}^{-1} \cr
C_{(6)}^T = - C_{(6)} & , & (\Gamma^{A})^T = C_{(6)} \Gamma^{A} C_{(6)}^{-1} \,,
\eea
giving the symmetry properties of the $\Gamma$-matrices. Similarly, complex conjugation of the $\Gamma$-matrices is given by
\bea
(\gamma^{a})^* & = & - B_{(5)} \gamma^{a} B_{(5)}^{-1} \cr
(\Gamma^{A})^* & = & - B_{(6)} \Gamma^{A} B_{(6)}^{-1} \,,
\eea
where we define the matrices $B_{(5)} = C_{(5)} \gamma^0$ and $B_{(6)} = C_{(6)} \Gamma^0$, satisfying $B B^* = - \id$.

We will next consider spinors $\Lambda^{\alpha}$, in either five or six dimensions, that carry an additional $\USp(4)$ spinor index in agreement with the application to $(2,0)$ theory. The conjugate spinor is defined as
\beq
\overline{\Lambda}{}^{\alpha} = (\Lambda^{\alpha})^T C
\eeq
and the charge conjugate spinor as
\beq
(\Lambda^{\alpha})^C = B^{-1} (\Lambda^{\alpha})^* \,.
\eeq
(With our conventions the Dirac conjugate is given by $\overline{(\Lambda^{\alpha})^C}$.) It is not possible in neither $4+1$ nor $5+1$ dimensions to define ordinary Majorana spinors due to the fact that $B B^* = -\id$. It is, however, possible to make use of the symplectic structure $M_{\alpha \beta}$ (described in detail in the previous subsection) to impose a consistent symplectic Majorana reality condition according to
\beq
\label{eqn:SymplecticMajoranaCondition}
(\Lambda^\alpha )^* = M_{\alpha \beta} B \Lambda^{\beta} \,.
\eeq
All the spinors we consider will satisfy this condition which implies that we have
\bea
(\Lambda^{\alpha})^C = M_{\alpha \beta} \Lambda^{\beta} = \Lambda_{\alpha} & , & \overline{(\Lambda^{\alpha})^C} = M_{\alpha \beta} \overline{\Lambda}{}^{\beta} = \overline{\Lambda}_{\alpha} 
\eea
for the charge conjugate spinors.

In the action functionals the spinors appear exclusively as $\USp(4)$ invariant bilinears. The symmetry and reality properties of such bilinears can be derived from the defining relations for the charge conjugation matrix and the properties of the symplectic metric $M_{\alpha \beta}$. For our purposes the relevant relations are
\beq
\label{eqn:SymmetryPropertiesFiveDimensions}
\overline{\psi}_{\alpha} \lambda^{\alpha} = - \overline{\lambda}_{\alpha} \psi^{\alpha} \,\,\, , \,\,\, \overline{\psi}_{\alpha} \gamma^{a} \lambda^{\alpha} = - \overline{\lambda}_{\alpha} \gamma^{a} \psi^{\alpha} \,\,\, , \,\,\, \overline{\psi}_{\alpha} \gamma^{ab} \lambda^{\alpha} = \overline{\lambda}_{\alpha} \gamma^{ab} \psi^{\alpha} 
\eeq
\beq
\label{eqn:RealityPropertiesFiveDimensions}
( \overline{\psi}_{\alpha} \lambda^{\alpha} )^* = \overline{\psi}_{\alpha} \lambda^{\alpha} \,\,\, , \,\,\, ( \overline{\psi}_{\alpha} \gamma^{a} \lambda^{\alpha} )^* = - \overline{\psi}_{\alpha} \gamma^{a} \lambda^{\alpha} \,\,\, , \,\,\, ( \overline{\psi}_{\alpha} \gamma^{ab} \lambda^{\alpha} )^* = \overline{\psi}_{\alpha} \gamma^{ab} \lambda^{\alpha}
\eeq
for spinors $\lambda^{\alpha}$ and $\psi^{\alpha}$ in $4+1$ dimensions and similarly
\beq
\label{eqn:SymmetryPropertiesSixDimensions}
\overline{\Psi}_{\alpha} \Lambda^{\alpha} = - \overline{\Lambda}_{\alpha} \Psi^{\alpha} \,\,\, , \,\,\, \overline{\Psi}_{\alpha} \Gamma^{A} \Lambda^{\alpha} = - \overline{\Lambda}_{\alpha} \Gamma^{A} \Psi^{\alpha} \,\,\, , \,\,\, \overline{\Psi}_{\alpha} \Gamma^{AB} \Lambda^{\alpha} = \overline{\Lambda}_{\alpha} \Gamma^{AB} \Psi^{\alpha} \eeq
\beq
\label{eqn:RealityPropertiesSixDimensions}
( \overline{\Psi}_{\alpha} \Lambda^{\alpha} )^* = \overline{\Psi}_{\alpha} \Lambda^{\alpha} \,\,\, , \,\,\, ( \overline{\Psi}_{\alpha} \Gamma^{A} \Lambda^{\alpha} )^* = - \overline{\Psi}_{\alpha} \Gamma^{A} \Lambda^{\alpha} \,\,\, , \,\,\, ( \overline{\Psi}_{\alpha} \Gamma^{AB} \Lambda^{\alpha} )^* = \overline{\Psi}_{\alpha} \Gamma^{AB} \Lambda^{\alpha}
\eeq
for spinors $\Lambda^{\alpha}$ and $\Psi^{\alpha}$ in $5+1$ dimensions.

In six dimensions the Dirac spinor representation is decomposed according to the eigenvalue of the chirality operator $\Gamma$, which we define to be
\beq
\Gamma = \Gamma^0 \Gamma^1 \ldots \Gamma^5\,.
\eeq
so that we can consider Weyl spinors of definite chirality
\beq
\label{eqn:WeylConditionSixDimensions}
\Gamma \Lambda^{\alpha} = \pm \Lambda^{\alpha} \,.
\eeq
The chirality condition is compatible with (\ref{eqn:SymplecticMajoranaCondition}) in six dimensions, admitting symplectic Majorana-Weyl spinors. Finally, we consider the matrices $\gamma^{0},\ldots,\gamma^{3}$ which generate the Clifford algebra in $3+1$ dimensions. Here we can define a chirality operator similar to the one in six dimensions, which provides the final generator
\beq
\gamma^{4} = \gamma = i\gamma^{0} \gamma^{1} \ldots \gamma^{3}
\eeq
of the Clifford algebra in $4+1$ dimensions.

\end{appendix}


\end{document}